\documentclass[letterpaper]{article}

\usepackage{aaai}
\nocopyright
\usepackage{times}
\usepackage{helvet}
\usepackage{courier}
\usepackage[T1]{fontenc}
\usepackage{url}
\usepackage{graphicx}
\graphicspath{{./}}
\usepackage{amsmath}
\usepackage{amsfonts}
\usepackage{amssymb}
\usepackage{dsfont}
\usepackage{booktabs}
\usepackage{multirow}
\usepackage{array}
\usepackage{adjustbox}
\usepackage{threeparttable}
\usepackage{diagbox}
\usepackage{subfigure}
\usepackage{placeins}
\usepackage{float}
\usepackage{xcolor}
\usepackage{microtype}

\DeclareMathOperator*{\argmin}{argmin}

\let\AAAIsection\section
\let\AAAIsubsection\subsection
\newcommand{\inputchapter}[3]{%
  \AAAIsection{#1}%
  \label{#2}%
  \begingroup
  \let\section\AAAIsubsection
  \input{#3}%
  \endgroup
}

\title{CITRUS: Candidate Inference and Temporal-tracking for Reliable, Unobtrusive Sensing of Wearable Heart Rate under Motion}

\author{
Yi Wang\\
Tsinghua University
}

\begin{document}

\maketitle

\begin{abstract}
Wearable photoplethysmography (PPG) provides continuous heart-rate measurements, but its accuracy degrades under motion. In the ring-platform benchmark, the best supervised baseline reaches 5.33~BPM mean absolute error (MAE) on the overall heart-rate task. In the motion-focused ring-only audit, a supervised LSTM baseline reaches $14.39 \pm 0.47$~BPM MAE on motion windows, and simple smoothing and ACC priors reduce this only to $13.00 \pm 0.41$~BPM. This thesis addresses motion-corrupted HR estimation through three connected stages: candidate-frequency identification, temporal decoding, and reliability-aware reporting.

The study first evaluates two wearable PPG ring variants against clinical references for heart rate, respiration, SpO$_2$, and blood pressure in 54 participants. The proposed system then uses two layers. The \textbf{estimation layer} converts each window into approximately 140 frequency candidates, scores candidates using agreement among independent estimators, and applies causal Viterbi decoding with a physiological transition penalty. The \textbf{reporting layer} estimates reliability and applies a learned accept/hold/reject policy. Reporting only the most-confident 50\% of motion windows reduces motion MAE from 10.8 to 6.2~BPM. The heart-rate estimator uses fewer than 100k parameters and runs within a microcontroller-class compute budget. Additional PPG-DaLiA experiments evaluate the same candidate-selection and temporal-decoding estimator on an independent wrist-BVP cohort.

\end{abstract}

  \AAAIsection{Introduction}  \label{chp:intro}  \begingroup
  \let\section\AAAIsubsection
  %% Chapter 1: Introduction
%% Focused single-topic thesis: motion-robust, deployable wearable-PPG heart rate.
%% Chapter title \chapter{Introduction} is declared in the master file (000_thesis.tex);
%% this file contains only \section-level content, matching the existing convention.

\section{The Problem: a Motion-Corrupted View of Heart Rate}
\label{sec:intro_problem}

Wearable photoplethysmography (PPG) enables low-cost and continuous cardiovascular monitoring~\cite{tang2025tau,fiore2024use,wang2025computing}. A PPG sensor illuminates the skin with light-emitting diodes and records intensity changes in the reflected light. These intensity changes contain a pulsatile component caused by blood-volume variation, from which heart rate can be estimated. The main limitation is motion sensitivity. Motion changes the sensor--skin contact and optical path, introducing artifacts that can overlap with the cardiac frequency band. At rest, ring PPG can track heart rate close to a clinical reference. During motion, the cardiac component overlaps with motion frequencies and harmonics, which increases estimation error. This thesis studies heart-rate estimation from ring PPG under these motion conditions.

\section{Progressive Decomposition of the Motion Problem}
\label{sec:intro_problem_decomposition}

The central problem is a single failure mode: motion corrupts the PPG signal and makes the cardiac component difficult to identify. The proposed system addresses this problem progressively. It first constructs plausible heart-rate candidates from each window, then scores candidates using local evidence, then enforces temporal consistency across windows, and finally decides whether the resulting estimate should be reported.

This progression follows the information available at each stage. Spectral ambiguity is local to a window. Candidate ambiguity requires comparing multiple estimators and harmonic hypotheses. Temporal inconsistency requires cross-window decoding. Deployment requires a reporting decision, because a wearable device may report a new value, hold the previous trusted value, or reject an unreliable window. The resulting decomposition is shown in Table~\ref{tab:intro_decomposition}.

\begin{table*}[t]
\centering
\caption{Progressive decomposition of the motion-robust heart-rate estimation problem.}
\label{tab:intro_decomposition}
\begin{adjustbox}{max width=\textwidth}
\begin{tabular}{lllll}
\toprule
\textbf{Layer} & \textbf{Difficulty} & \textbf{Nature} & \textbf{Solution} & \textbf{Chapter} \\
\midrule
Estimation & Spectral ambiguity (motion artifacts) & single-window, frequency & Candidate generation & Ch.~4 \\
Estimation & Candidate ambiguity (which is true) & single-window, value & Candidate scoring & Ch.~4 \\
Estimation & Temporal inconsistency (cross-window jumps) & global, temporal & Causal Viterbi & Ch.~4 \\
Reporting & Deployment decision (emit / hold / reject) & global, decision & Reliability + policy & Ch.~5 \\
\bottomrule
\end{tabular}
\end{adjustbox}
\end{table*}

This decomposition defines the algorithmic sequence used in this thesis: local candidate construction, candidate scoring, temporal decoding, and reliability-aware reporting.

\section{Approach}
\label{sec:intro_approach}

The work has four parts. First, we use a $\tau$-Ring-based wearable PPG platform to establish a 54-participant multi-vital benchmark against clinical references, quantifying the error profile across vital-sign tasks and motion scenarios. Second, we develop an estimation layer based on candidate selection and causal Viterbi decoding. Third, we add a reporting layer consisting of reliability scoring and a learned selective-reporting policy, and evaluate the compute budget for microcontroller deployment. Fourth, we evaluate the estimator on PPG-DaLiA, an independent wrist-BVP dataset, with DaLiA-specific external evaluation. We refer to the resulting two-layer system as \textbf{CITRUS} (\textbf{C}andidate \textbf{I}nference and \textbf{T}emporal-tracking for \textbf{R}eliable, \textbf{U}nobtrusive \textbf{S}ensing). Component ablations are used to separate the contribution of candidate generation, temporal decoding, reliability scoring, and reporting policy.

\section{Contributions}
\label{sec:intro_contributions}

The main contributions of this thesis are:

\begin{itemize}
    \item \textbf{A $\tau$-Ring-based multi-vital benchmark} against clinical references (54 participants), quantifying error profiles across vital-sign tasks and motion scenarios.
    \item \textbf{A candidate-based HR estimation layer} that combines DSP-generated HR candidates, learned candidate scoring, and causal Viterbi decoding.
    \item \textbf{A reliability-aware reporting layer} with reliability scoring and a learned accept/hold/reject policy, reducing motion-window MAE from 10.8 to 6.2~BPM at 50\% coverage under an estimated MCU-level compute budget.
    \item \textbf{A PPG-DaLiA evaluation} on an independent wrist-BVP dataset, analyzing DaLiA-specific training and cross-dataset behavior of the candidate-selection and temporal-decoding estimator.
\end{itemize}

\section{Thesis Organization}
\label{sec:intro_organization}

Chapter~2 reviews related work on wearable HR, temporal decoding, and selective prediction. Chapter~3 establishes the platform and benchmark. Chapter~4 develops the estimation layer. Chapter~5 develops the reporting layer and demonstrates MCU deployment. Chapter~6 reports external replication and concludes.
  \endgroup

  \AAAIsection{Literature Review}  \label{chp:literature}  \begingroup
  \let\section\AAAIsubsection
  %% Chapter 2: Literature Review
%% Re-scoped to the HR-under-motion topic. All cite keys are defined across
%% references.bib and references_extra.bib (verified: 0 undefined). The classical
%% Viterbi/HMM decoding and selective-prediction/calibration references are supplied
%% in references_extra.bib with canonical bibliographic metadata.

\section{Wearable PPG Heart-Rate Estimation and Motion Artifacts}
\label{sec:lit_wearable}

Photoplethysmography (PPG) measures the blood-volume pulse from skin-surface optics, and is now embedded in watches, earbuds, and rings. Ring-worn PPG is attractive because the finger's thin skin and dense capillary bed yield a strong pulse, and a growing body of work uses smart rings for continuous physiological monitoring~\cite{tang2025tau,fiore2024use,wang2025computing}, including atrial-fibrillation screening~\cite{kwon2020detection}, oxygen saturation under hypoxia~\cite{mastrototaro2024performance}, and cuffless blood pressure~\cite{kim2023firstinhuman}.

The dominant obstacle is motion. Mechanical coupling between the optical path and limb movement injects artifacts whose frequencies overlap---and often dominate---the cardiac band; during exercise the motion spectrum can be stronger than the pulse itself~\cite{zhang2015troika,jarchi2017description}. Mitigations range from adaptive filtering against an auxiliary motion reference such as an accelerometer or gyroscope~\cite{lee2019motion} to dataset-level efforts that capture PPG with synchronized inertial signals for benchmarking under realistic activity~\cite{reiss2019deep,schmidt2018introducing,meierwildppg}. Even with such references, signal quality varies widely across devices and protocols~\cite{wolling2020quest,ho2025wfppg}.

\section{Heart-Rate Estimation Methods}
\label{sec:lit_methods}

Classical HR estimation works in the time or frequency domain. Time-domain methods detect individual beats and convert inter-beat intervals to rate~\cite{vangent2019heartpy,gent2019analysing}; frequency-domain methods locate the dominant spectral peak. Under motion, neither is reliable alone, which motivates frameworks that combine signal decomposition, spectral peak selection, and temporal verification. A representative example is TROIKA, which couples singular-spectrum decomposition with sparse spectral reconstruction and peak tracking for wrist PPG during intensive exercise~\cite{zhang2015troika}. Learning-based methods instead regress HR end-to-end; Deep~PPG showed that convolutional networks trained on large activity datasets improve motion robustness over hand-tuned pipelines~\cite{reiss2019deep}. Under motion artifacts, the main difficulty is not detecting a single spectral peak, but discriminating among multiple plausible candidates. These candidates may correspond to the true pulse, motion components, or harmonic multiples. Classical spectral trackers usually rely on hand-designed rules for peak selection; end-to-end models do not achieve the best motion performance on this dataset, and their errors are harder to attribute. This thesis therefore adopts candidate selection as the main estimation formulation. Multiple DSP estimators first generate HR candidates, and a learned scorer ranks them using source method, channel, spectral strength, harmonic relation, consensus, and temporal-distance features, making motion-induced candidate disambiguation explicit.

\section{Temporal Decoding}
\label{sec:lit_decoding}

A per-window estimate ignores a strong prior: heart rate changes slowly. Spectral-tracking methods such as TROIKA already exploit this by constraining each estimate to lie near the previous one and verifying it over time~\cite{zhang2015troika}. More generally, choosing one value per window from a set of candidates, subject to a smoothness penalty between neighbours, is a shortest-path problem solved exactly by dynamic programming---the Viterbi algorithm~\cite{viterbi1967error}, classically used for sequence decoding in hidden Markov models~\cite{rabiner1989tutorial}. Chapter~\ref{chp:estimation} adopts a causal variant, so that decoding depends only on past windows and remains deployable in real time.

\section{Selective Prediction and Reliability Estimation}
\label{sec:lit_selective}

Selective prediction studies when a model should return an output. In classification or regression with a reject option, the model may withhold predictions for low-confidence inputs, forming a coverage--error trade-off~\cite{elyaniv2010foundations,geifman2017selective}. Most implementations use a confidence score, and the decision threshold determines the coverage--error trade-off.

In wearable HR estimation, this threshold should not be assumed to be device-independent. PPG quality, motion artifacts, candidate distributions, and model-score distributions can vary across devices and cohorts. A reporting threshold calibrated on one sensing setting may therefore not transfer directly to another. This thesis treats reliability estimation and reporting policy as calibration problems tied to the target sensing setting, rather than using a fixed universal threshold.

This thesis applies selective prediction in the reporting layer. A reliability model scores each HR estimate. A policy module then decides whether to report the current value, hold the previous trusted value, or reject the window.

\section{On-Device and Efficient Inference}
\label{sec:lit_ondevice}

Wearables run on microcontroller-class hardware with kilobytes of memory and tight power budgets, so model size and compute matter as much as accuracy. The TinyML literature targets this regime directly~\cite{Ji2023TinyML,coffen2021tinydl}, through compact architectures~\cite{tan2020efficientnet} and hardware--software co-design; at the extreme, micro-power optical front-ends bring beat detection into the sensor itself~\cite{boukhayma2021ring}. This thesis reports parameter, memory, and FLOP budgets for the full pipeline (Chapter~\ref{chp:reporting}) and shows it fits a Cortex-M-class device.
  \endgroup

  \AAAIsection{A Wearable PPG Platform and Multi-Vital Benchmark}  \label{chp:platform}  \begingroup
  \let\section\AAAIsubsection
  %% Chapter 3: A Ring PPG Platform and Multi-Vital Benchmark
%% Establishes the platform, quantifies the wearable-vs-lab gap across vitals,
%% and identifies motion as the main failure mode, motivating Chapters 4-5.
%% Floats migrated verbatim from the original 03_methods.tex / 04_experiments.tex.
%% Figure assets required in the build dir: ring_demonstrate.pdf.
%% (tab:healthring_stats wrapped in threeparttable to fix the original's bare \tablenotes.)

\section{Hardware Platform}
\label{sec:plat_hardware}

The hardware platform follows the $\tau$-Ring cardiovascular sensing platform~\cite{tang2025tau}. \textbf{Ring~1} denotes the reflective design, where the emitter and photodetector are placed on the same side of the finger and the GH3026 sensor measures reflected light. \textbf{Ring~2} denotes the transmissive design, where light passes through finger tissue and is measured by an AFE4403 front end. Both variants use an nRF52840 controller and a QMA6100P 3-axis accelerometer, and both record IR and red PPG at 100~Hz (Figure~\ref{fig:ring_demonstrate}).

\begin{figure}[htbp]
    \centering
    \includegraphics[width=0.9\columnwidth]{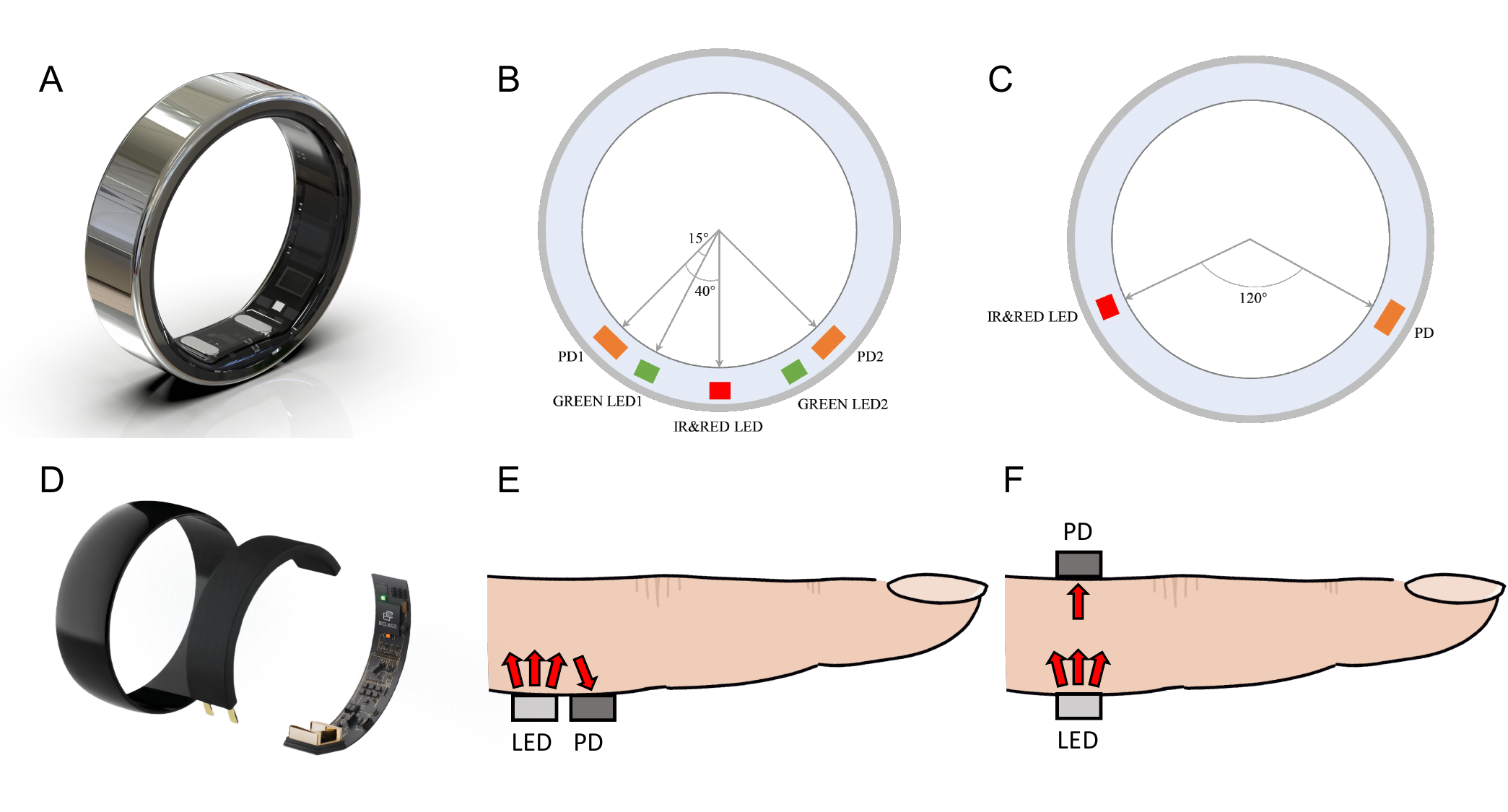}
    \caption{\textbf{Ring design and sensor layout.} (A) Physical structure of the ring. (D) Internal components including battery and chip. (B) and (E) Sensor configuration for Ring~1 (reflective). (C) and (F) Sensor configuration for Ring~2 (transmissive).}
    \label{fig:ring_demonstrate}
\end{figure}

\section{Dataset and References}
\label{sec:plat_dataset}

The benchmark used in this thesis contains 54 participants and 49.76~h of recordings across three cohorts: Study~1 (stimulus-evoked states), Study~2 (daily activities), and Study~3 (treadmill running). The recordings are paired with clinical references: a finger oximeter, a respiratory band, a blood-pressure monitor, and an FDA-cleared VivaLNK ECG patch for treadmill sessions (Table~\ref{tab:healthring_stats}).

\begin{table*}[t]
\centering
\caption{Ring-platform benchmark statistics across three study cohorts.}
\label{tab:healthring_stats}
\begin{adjustbox}{max width=\textwidth}
\begin{threeparttable}
\small
\begin{tabular}{lccccc}
\hline
\textbf{Cohort} & \textbf{Subjects} & \textbf{Duration (h)} & \textbf{Rings} & \textbf{Labels} & \textbf{Reference} \\
\hline
Study~1 (Stimulus) & 26 & 16.31 & Ring~1, Ring~2 & HR, RR, SpO$_2$, BP & Oximeter, Resp Band, BP Monitor \\
Study~2 (Daily) & 16 & 11.90 & Ring~1, Ring~2 & HR, RR & Oximeter, Resp Band \\
Study~3 (Treadmill) & 20 & 21.55 & Ring~1 only & HR, RR & VivaLNK ECG Patch \\
\hline
Total & 54$^*$ & 49.76 & --- & --- & --- \\
\hline
\end{tabular}
\begin{tablenotes}
\footnotesize
\item $^*$8 participants completed both Study~1 and Study~2 (42 sessions total across Studies~1 and~2).
\end{tablenotes}
\end{threeparttable}
\end{adjustbox}
\end{table*}

\section{Benchmark and Multi-Vital Results}
\label{sec:plat_benchmark}

Table~\ref{tab:summary_all_thesis} summarizes the non-treadmill multi-vital benchmark. Ring~1 and Ring~2 refer to the reflective and transmissive prototypes defined in Section~\ref{sec:plat_hardware}. The table reports only the lowest MAE for each vital-sign target and ring. Ring~1 gives lower error for HR, RR, SBP, and DBP, while Ring~2 gives lower SpO$_2$ error. The motion-focused HR experiments below use the Ring~1 HR subset with a stricter ring-only input boundary.

\begin{table*}[t]
\centering
\caption{Non-treadmill multi-vital benchmark summary. Values are best MAE for each task and ring.}
\label{tab:summary_all_thesis}
\begin{adjustbox}{max width=\textwidth}
\begin{tabular}{llccc}
\toprule
\textbf{Task} & \textbf{Unit} & \textbf{Ring~1 (reflective)} & \textbf{Ring~2 (transmissive)} & \textbf{Lower MAE} \\
\midrule
HR & BPM & $5.33 \pm 0.17$ & $7.32 \pm 0.17$ & Ring~1 \\
RR & breaths/min & $2.98 \pm 0.06$ & $3.28 \pm 0.07$ & Ring~1 \\
SpO$_2$ & \% & $1.72 \pm 0.04$ & $1.48 \pm 0.04$ & Ring~2 \\
SBP & mmHg & $12.98 \pm 0.74$ & $14.81 \pm 1.07$ & Ring~1 \\
DBP & mmHg & $7.64 \pm 0.51$ & $8.11 \pm 0.72$ & Ring~1 \\
\bottomrule
\end{tabular}
\end{adjustbox}
\end{table*}

\section{Heart-Rate Error under Motion}
\label{sec:plat_motion}

The motion-focused HR analysis uses 31 Ring~1 subject files from RingDatasetV2, after excluding synchronization windows and windows without HR labels. The resulting split contains 2009 windows: 1742 static windows and 267 motion windows from deep-squat, striding, and shaking-head protocols. Runtime inputs are restricted to ring PPG, ring ACC, and ring quality features; reference-device variables and $y_t$ are not model inputs.

\begin{table*}[t]
\centering
\caption{Ring-only HR error stratified by motion. Values are MAE in BPM over five seeds.}
\label{tab:ring_motion_baseline}
\begin{adjustbox}{max width=\textwidth}
\begin{tabular}{lccc}
\toprule
\textbf{Configuration} & \textbf{Static MAE} & \textbf{Motion MAE} & \textbf{Overall MAE} \\
\midrule
LSTM baseline & $5.52 \pm 0.33$ & $14.39 \pm 0.47$ & $6.70 \pm 0.27$ \\
LSTM with smoothing and ACC priors & $5.33 \pm 0.29$ & $13.00 \pm 0.41$ & $6.35 \pm 0.30$ \\
HGB feature probe & $3.85 \pm 0.00$ & $13.08 \pm 0.00$ & $5.08 \pm 0.00$ \\
Candidate scorer + causal Viterbi & $2.63 \pm 0.06$ & $10.91 \pm 0.73$ & $3.73 \pm 0.13$ \\
\bottomrule
\end{tabular}
\end{adjustbox}
\end{table*}

Motion is the dominant error source under this deployable-input protocol. A supervised LSTM reaches $14.39 \pm 0.47$~BPM motion MAE, and adding simple smoothing and ACC-derived priors reduces this only to $13.00 \pm 0.41$~BPM. The candidate-based estimator developed in Section~\ref{sec:est_viterbi} lowers full-coverage motion MAE to $10.91 \pm 0.73$~BPM, but motion windows remain substantially harder than static windows. This motivates the reporting layer in Section~\ref{sec:rep_reliability}.
  \endgroup

  \AAAIsection{The Estimation Layer: Candidate Selection and Causal Viterbi}  \label{chp:estimation}  \begingroup
  \let\section\AAAIsubsection
  %% Chapter 4: The Estimation Layer — Candidate Selection and Causal Viterbi
%% Produces an accurate internal HR under motion (difficulties 1-3 of Sec. intro_two_questions).

This chapter defines the estimation layer for heart-rate estimation under motion (the first three stages in Section~\ref{sec:intro_problem_decomposition}). Let $t$ index analysis windows, $y_t$ denote the reference HR used for training and evaluation, and $\hat{y}_t$ denote the deployed estimate. The estimator observes only ring PPG/ACC-derived features at runtime.

\section{Candidate Generation}
\label{sec:est_candidates}

Each short analysis window is converted into candidate HR values using Welch spectral peaks, autocorrelation peaks, and peak-interval estimates. The estimators are applied to both PPG channels (IR/red). Harmonic variants are added and candidates outside the physiological range are removed:

\begin{equation}
\begin{aligned}
\mathcal{C}_t =
\{\, \alpha b :\;& b \in \mathcal{B}_{t,q,m},
q \in \{\mathrm{IR},\mathrm{red}\},\\
& m \in \mathcal{M},
\alpha \in \{0.5,1,2\}\,\}
\cap [35,220].
\end{aligned}
\label{eq:candidate_set}
\end{equation}
where $\mathcal{B}_{t,q,m}$ is the set of BPM values proposed by estimator $m$ on channel $q$. The resulting candidate set is written as $\mathcal{C}_t=\{c_{t,k}\}_{k=1}^{K_t}$ with a validity mask for padded candidates. It contains approximately 140 valid candidates per window (median 140, range 57--160). The accelerometer is used before candidate extraction to select motion-quiet segments; it is not a candidate source.

Candidate-set coverage is measured by the best available candidate in each window:
\begin{equation}
e_t^{\mathrm{cov}} = \min_{c \in \mathcal{C}_t} |c-y_t|,
\qquad
\mathrm{MAE}_{\mathrm{cov}} = \frac{1}{|\mathcal{T}|}
\sum_{t \in \mathcal{T}} e_t^{\mathrm{cov}} .
\label{eq:candidate_coverage}
\end{equation}
The DSP candidate set has a coverage MAE of 0.6~BPM, but candidate density affects this number. Under the same HR range, shuffled references remain near 1.2~BPM, and a 160-state uniform grid reaches 0.32~BPM. This metric measures candidate coverage, not deployable selection quality.

\section{Candidate-Source Control}
\label{sec:est_candidate_source_control}

To separate candidate coverage from candidate source, the DSP candidates are replaced with a 160-state uniform grid over 35--220~BPM:
\begin{equation}
\mathcal{G} =
\left\{35 + (k-1)\frac{220-35}{159}\right\}_{k=1}^{160}.
\label{eq:uniform_grid}
\end{equation}
The scorer, leave-one-subject-out split, five seeds, and causal Viterbi decoder are kept fixed. The uniform-grid candidates contain only candidate BPM and constant grid metadata; they remove DSP source, channel, spectral quality, consensus, and harmonic features.

\begin{table*}[t]
\centering
\small
\caption{Candidate-source control under the same scorer and causal decoder. MAE is reported in BPM over five seeds.}
\label{tab:candidate_source_control}
\begin{adjustbox}{max width=\textwidth}
\begin{tabular}{lllll}
\toprule
Candidate source & Raw input & Motion & Static & Overall \\
\midrule
DSP candidates & IR/red/ACC & $10.91 \pm 0.73$ & $2.63 \pm 0.06$ & $3.73 \pm 0.13$ \\
DSP candidates & IR/red & $10.49 \pm 0.84$ & $2.56 \pm 0.10$ & $3.61 \pm 0.19$ \\
Uniform grid & IR/red/ACC & $12.06 \pm 0.27$ & $4.60 \pm 0.17$ & $5.60 \pm 0.14$ \\
Uniform grid & IR/red & $12.42 \pm 0.17$ & $4.32 \pm 0.33$ & $5.40 \pm 0.30$ \\
Uniform grid & ACC & $13.96 \pm 0.35$ & $10.83 \pm 0.23$ & $11.25 \pm 0.22$ \\
Uniform grid & none & $13.17 \pm 0.07$ & $9.57 \pm 0.08$ & $10.05 \pm 0.08$ \\
\bottomrule
\end{tabular}
\end{adjustbox}
\end{table*}

The uniform grid gives a lower coverage MAE than the DSP candidate set, but worse deployable estimates. With IR/red/ACC input, replacing DSP candidates with the grid increases motion MAE from $10.91$ to $12.06$~BPM and overall MAE from $3.73$ to $5.60$~BPM. With IR/red only, motion MAE increases from $10.49$ to $12.42$~BPM. The no-signal grid reaches $13.17$~BPM on motion but has poor static and overall accuracy. The ACC-only grid is worse than the PPG-based grid controls. These results indicate that dense HR coverage and temporal continuity explain part of the motion result, but the DSP candidate set provides signal-specific structure used by the scorer and decoder.

\section{Candidate Scoring, and the Redundant Encoder}
\label{sec:est_scoring}

Each candidate is represented by features $\phi_{t,k}$: source method, source channel, spectral power, consensus (nearby candidates or methods), distance to the previous window, harmonic status, and optional raw-window embedding. A small MLP maps these features to candidate logits,
\begin{equation}
\ell_{t,k} = f_\theta(\phi_{t,k}),
\qquad
p_{t,k} =
\frac{\exp(\ell_{t,k})}
{\sum_{j:m_{t,j}=1}\exp(\ell_{t,j})},
\label{eq:candidate_softmax}
\end{equation}
where $m_{t,k}$ is the validity mask. Training uses a soft target centered on the reference HR,

\begin{equation}
\begin{aligned}
q_{t,k} &=
\frac{m_{t,k}\exp(-|c_{t,k}-y_t|/\tau)}
{\sum_j m_{t,j}\exp(-|c_{t,j}-y_t|/\tau)},\\
\hat{y}^{\mathrm{row}}_t &= \sum_k p_{t,k} c_{t,k}.
\end{aligned}
\label{eq:candidate_target}
\end{equation}
The scorer is optimized with a cross-entropy term against $q_{t,k}$ and a Smooth-L1 term on $\hat{y}^{\mathrm{row}}_t$:
\begin{equation}
\mathcal{L}_{\mathrm{score}} =
-\lambda_{\mathrm{ce}}\sum_k q_{t,k}\log p_{t,k}
+\lambda_{\mathrm{reg}}\operatorname{Huber}_{\beta}
(\hat{y}^{\mathrm{row}}_t-y_t).
\label{eq:scorer_loss}
\end{equation}
This formulation ranks candidates using both reference proximity during training and candidate evidence at runtime. For example, a cluster near 72~BPM can be ranked above an isolated 143.5~BPM peak that is likely a $2\times$ harmonic.

Across five random seeds, removing the raw-waveform CNN encoder does not reduce accuracy. Overall MAE changes from $4.16 \pm 0.04$ to $4.15 \pm 0.23$~BPM ($-0.01$~BPM), and motion MAE changes from $11.30$ to $11.10$~BPM ($-0.20$~BPM). The result shows no measurable benefit from the raw-waveform encoder under the evaluated candidate-feature configuration. Removing it reduces the model size from 85k to 28k parameters.

\section{Stage-Specific Role of the Accelerometer}
\label{sec:est_acc}

A $2\times2$ ablation reports motion MAE over five seeds: full model \textbf{10.73}, dropping motion features 12.33, using uniform time slices without ACC segment selection 11.86, and removing ACC entirely 11.76. ACC therefore contributes a 1--1.6~BPM motion-MAE reduction during candidate generation. In the reliability model (Section~\ref{sec:rep_reliability}), by contrast, ACC features do not improve performance. In this pipeline, ACC is useful for stable-segment selection but not for downstream reliability scoring.

\section{Causal Viterbi Decoding}
\label{sec:est_viterbi}

Independent per-window scoring can produce physiologically implausible jumps. The decoder selects a minimum-cost path through the candidate graph:

\begin{equation}
\begin{aligned}
k_{1:T}^{*} &=
\argmin_{k_1,\ldots,k_T}
\sum_{t=1}^{T} E_t(k_t)
+\sum_{t=2}^{T}\Omega_t(k_{t-1},k_t),\\
\hat{y}_t &= c_{t,k_t^{*}}.
\end{aligned}
\label{eq:viterbi_objective}
\end{equation}
with emission cost $E_t(k)=w_E[-\log p_{t,k}]$ and transition cost

\begin{equation}
\resizebox{\columnwidth}{!}{$\displaystyle
\begin{aligned}
\Delta_{t,j,k} &= |c_{t,k}-c_{t-1,j}|,\\
\sigma_{t,j,k} &=
\operatorname{clip}\left(
 s_0
 \sqrt{\operatorname{clip}\left(
 \frac{(c_{t-1,j}+c_{t,k})/2}{h_0},
 a_{\min},a_{\max}\right)}
 (1+\eta_p p_{t,k}),
 \sigma_{\min},\sigma_{\max}\right),\\
d_{t,j,k} &= \max(\sigma_{t,j,k}, d_{\max}a_{\mathrm{step}}),\\
\Omega_t(j,k) &=
\lambda_{\mathrm{tr}}
\min\left[
\left(\frac{\Delta_{t,j,k}}{\sigma_{t,j,k}}\right)^2,
\Omega_{\max}
\right]
+\lambda_{\mathrm{jump}}
\left(\frac{\max(0,\Delta_{t,j,k}-d_{t,j,k})}{d_{t,j,k}}\right)^2 .
\end{aligned}$}
\label{eq:viterbi_transition}
\end{equation}
Here $s_0$ is the transition-scale parameter, $h_0$ is the reference HR level, and $p_{t,k}$ is the current-window candidate probability. The experiments use $s_0=10$, $h_0=80$, $a_{\min}=0.75$, $a_{\max}=1.8$, $\sigma_{\min}=6$, $\sigma_{\max}=30$, $\Omega_{\max}=9$, $d_{\max}=25$, $a_{\mathrm{step}}=0.8$, and $\lambda_{\mathrm{jump}}=0.25$; $\lambda_{\mathrm{tr}}$ is selected on the training fold. Dynamic programming computes the recursion
\begin{equation}
D_t(k) = E_t(k) +
\min_j \left[D_{t-1}(j)+\Omega_t(j,k)\right].
\label{eq:viterbi_recursion}
\end{equation}
The causal version applies this recursion without future windows. This is path selection over candidates rather than moving-average smoothing. If the highest-scored candidate in one window is 144~BPM, the decoded path may still select 72~BPM when the transition penalty is lower.

Sequence decoding improves MAE relative to independent per-window scoring over five seeds: overall $4.37 \rightarrow 3.73$, motion $12.94 \rightarrow 10.91$, and static $3.06 \rightarrow 2.63$. Causal decoding ($L=0$, past-only, deployable) gives 10.66~BPM motion MAE, compared with 10.88~BPM for offline full-sequence decoding. The causal version therefore preserves the observed benefit without using future windows. Motion errors under this scheme are bidirectional scatter (signed bias $-1.3$~BPM; harmonic collapse accounts for 2--3\% of errors), rather than a systematic pull toward the mean or harmonic lock-in.
  \endgroup

  \AAAIsection{The Reporting Layer: Reliability, Selective Policy, and Deployment}  \label{chp:reporting}  \begingroup
  \let\section\AAAIsubsection
  %% Chapter 5: The Reporting Layer — Reliability, Selective Policy, and Deployment
%% Decides whether the internal estimate is emitted (difficulty 4), and shows MCU deployability.

This chapter develops the reporting layer, which decides whether the internal estimate is emitted (the final stage in Section~\ref{sec:intro_problem_decomposition}). It also reports the compute and memory budget for microcontroller deployment. Results use the same 31-subject subset; the policy experiment in Section~\ref{sec:rep_policy} uses three seeds. As in Chapter~\ref{chp:estimation}, $y_t$ denotes the reference HR label for window $t$, and is used only for training and offline evaluation.

\section{Reliability Scoring}
\label{sec:rep_reliability}

The estimation layer always produces a value. Under heavy motion, it can still choose from a candidate set in which all candidates are unreliable. A histogram-based gradient-boosted tree (HGB) estimates absolute reliability, distinct from the relative candidate ranking in Section~\ref{sec:est_scoring}. For an estimate $\hat{y}_t$, the training error and continuous reliability target are
\begin{equation}
e_t = |\hat{y}_t-y_t|,
\qquad
r_t = \exp(-e_t/\tau_r).
\label{eq:reliability_target}
\end{equation}
The model predicts $\rho_t=g(\psi_t)\in[0,1]$, where $\psi_t$ includes PPG quality, inter-estimator disagreement, candidate dispersion, and motion intensity (44--57 features). The diagnostic AUC uses the high-error event
\begin{equation}
z_t^{(10)} = \mathds{1}[e_t>10~\mathrm{BPM}],
\label{eq:high_error_event}
\end{equation}
which measures whether $\rho_t$ ranks unreliable windows below reliable ones. The PPG-only reliability model reaches \textbf{AUC(err${>}10$) $=0.898$} over five seeds, with $r(\text{conf}, -\text{err}) = 0.539$. A modality ablation gives PPG-only \textbf{0.898}, PPG+ACC 0.888, and ACC-only 0.710, indicating that ACC features do not improve this reliability model (cf.\ Section~\ref{sec:est_acc}).

\section{Selective Reporting}
\label{sec:rep_selective}

Selective reporting keeps a subset $S_\alpha$ of windows ranked by reliability. For a target coverage $\alpha$,
\begin{equation}
S_\alpha = \{t:\rho_t \ge q_{1-\alpha}(\rho)\},
\qquad
\mathrm{Coverage}(S_\alpha)=\frac{|S_\alpha|}{|\mathcal{T}|},
\label{eq:coverage_set}
\end{equation}
and the reported error is
\begin{equation}
\mathrm{MAE}(S_\alpha)=
\frac{1}{|S_\alpha|}\sum_{t\in S_\alpha}|\hat{y}_t-y_t|.
\label{eq:selective_mae}
\end{equation}
Withholding the least-reliable windows lowers motion MAE monotonically: at coverage $1.00$, motion MAE is $\approx 10.8$~BPM; at $0.80$, it is $9.35$~BPM; and at $0.50$, it is \textbf{6.2}~BPM.

\section{Finite-State Reporting Policy}
\label{sec:rep_policy}

The reporting layer chooses among three actions: accept the current estimate, hold the last trusted estimate, or reject the current window. The policy state is
\begin{equation}
s_t=(b_t,h_t),
\qquad
a_t\in\mathcal{A}=\{\mathrm{accept},\mathrm{hold},\mathrm{reject}\},
\label{eq:policy_state_action}
\end{equation}
where $b_t$ is the reliability bin of $\rho_t$ and $h_t$ is the hold count since the last accepted estimate. Let $\tilde{y}_{t-1}$ be the last trusted HR value. The per-window cost used for policy learning is
\begin{equation}
\ell_t(a_t)=
\begin{cases}
|\hat{y}_t-y_t|, & a_t=\mathrm{accept},\\
|\tilde{y}_{t-1}-y_t| + \beta h_t, & a_t=\mathrm{hold},\\
\lambda_{\mathrm{rej}}, & a_t=\mathrm{reject}.
\end{cases}
\label{eq:policy_cost}
\end{equation}
The hold term penalizes stale values, and $\lambda_{\mathrm{rej}}$ controls the coverage--error trade-off. Value iteration solves the finite-state dynamic program
\begin{equation}
\begin{aligned}
Q(s,a) &= \ell(s,a)+\gamma\sum_{s'}P(s'|s,a)V(s'),\\
V(s) &= \min_{a\in\mathcal{A}} Q(s,a),\\
\pi^*(s) &= \argmin_{a\in\mathcal{A}} Q(s,a).
\end{aligned}
\label{eq:value_iteration}
\end{equation}
At inference time, $y_t$ is unavailable; the learned table $\pi^*(s_t)$ uses only reliability bin and hold state. Unlike a fixed threshold, the policy conditions on how long the system has been holding the previous value. Table~\ref{tab:policy_ring_curve} reports the learned-policy coverage--error curve for the ring dataset. The rows are indexed by the rejection cost $\lambda_{\mathrm{rej}}$; the resulting motion coverage is therefore an output of the policy, not a preset coverage grid.

\begin{table*}[t]
\centering
\caption{Finite-state reporting policy on the ring dataset. MAE is reported in BPM for the emitted output stream after accept/hold/reject decisions; it is not the raw estimator MAE reported in Table~\ref{tab:candidate_source_control}. Coverage is the fraction of windows reported by the policy.}
\label{tab:policy_ring_curve}
\begin{adjustbox}{max width=\textwidth}
\begin{tabular}{rrrrr}
\toprule
$\lambda_{\mathrm{rej}}$ & Overall coverage & Overall MAE & Motion coverage & Motion MAE \\
\midrule
30 & 1.00 & 3.97 & 1.00 & 9.99 \\
18 & 0.99 & 3.78 & 0.94 & 9.50 \\
12 & 0.99 & 3.78 & 0.94 & 9.50 \\
8  & 0.97 & 3.61 & 0.87 & 9.23 \\
6  & 0.74 & 2.44 & 0.36 & 3.24 \\
4  & 0.65 & 2.26 & 0.31 & 2.90 \\
2  & 0.14 & 1.70 & 0.09 & 1.65 \\
\bottomrule
\end{tabular}
\end{adjustbox}
\end{table*}

The full-coverage estimator baseline remains the causal Viterbi result in Table~\ref{tab:candidate_source_control}: $10.91 \pm 0.73$~BPM motion MAE over five seeds. Table~\ref{tab:policy_ring_curve} instead evaluates the emitted reporting stream. At $\lambda_{\mathrm{rej}}=30$, the policy reports every motion window, but it may emit a held value rather than the current estimator output; this three-seed policy run gives 9.99~BPM motion MAE. At $\lambda_{\mathrm{rej}}=4$, the policy reports 31\% of motion windows and the reported-window motion MAE is 2.90~BPM. Low-coverage rows should therefore be interpreted as reporting-output operating points, not as replacement full-coverage estimator accuracy.

\section{Ablation Summary}
\label{sec:rep_ablation}

\begin{table*}[t]
\centering
\caption{Ablation summary for the motion-robust HR system.}
\label{tab:reporting_ablation}
\begin{adjustbox}{max width=\textwidth}
\begin{tabular}{llll}
\toprule
\textbf{Module} & \textbf{Ablation} & \textbf{Conclusion} & \textbf{Key number} \\
\midrule
Scoring & disable raw-waveform CNN & encoder redundant & $\Delta$overall $-0.01$~BPM (5 seed) \\
Candidate gen.\ & ACC segment / motion features & needed upstream & 10.73 vs.\ 11.76 motion \\
Reliability & feature modality & PPG-driven, ACC redundant & AUC 0.898 vs.\ 0.888 \\
Viterbi & latency $L{=}0$..offline & causal $\approx$ offline & 10.66 vs.\ 10.88 motion \\
Policy & learned vs.\ threshold & gains at high coverage & $-1.2$~BPM @\,cov${=}1.0$ \\
\bottomrule
\end{tabular}
\end{adjustbox}
\end{table*}

\section{Deployment}
\label{sec:rep_deployment}

The HR model has 85k parameters (0.34~MB fp32 / 83~KB int8), or \textbf{28k} parameters without the raw-waveform encoder. It requires ${\sim}18$--$19$~MFLOPs per window. The full stack is ${\sim}1.45$~MB fp32, dominated by the 1.1~MB HGB model, which can be compressed further. The estimated runtime is ${\approx}0.2$~s per window on a Cortex-M4F and ${<}1$~ms per window on an NPU-equipped SoC. These estimates fit a microcontroller-class real-time deployment budget.
  \endgroup

  \AAAIsection{External Replication and Conclusion}  \label{chp:conclusion}  \begingroup
  \let\section\AAAIsubsection
  %% Chapter 6: External Replication and Conclusion

\section{PPG-DaLiA External Evaluation}
\label{sec:concl_dalia}

PPG-DaLiA is used as an external wrist-BVP evaluation cohort. The dataset contains 15 subjects with single-channel wrist BVP and 3-axis accelerometry. The estimator is trained and evaluated with subject-disjoint LOSO on DaLiA; no ring-trained weights are used. Table~\ref{tab:dalia_external} reports the full-coverage estimator and selected reporting operating points. Table~\ref{tab:dalia_policy_curve} reports the finite-state policy curve on DaLiA.

\begin{table}[H]
\centering
\caption{PPG-DaLiA external evaluation. MAE is reported in BPM.}
\label{tab:dalia_external}
\begin{adjustbox}{max width=\columnwidth}
\begin{tabular}{lccc}
\toprule
\textbf{Configuration} & \textbf{Coverage} & \textbf{Overall MAE} & \textbf{Motion MAE} \\
\midrule
Row-wise candidate scorer & 1.00 & $8.70 \pm 0.21$ & $14.22 \pm 0.17$ \\
Causal sequence decoding & 1.00 & $7.63 \pm 0.21$ & $12.62 \pm 0.28$ \\
Fixed-threshold selective reporting & 0.50 & $3.12 \pm 0.02$ & $7.30 \pm 0.51$ \\
Finite-state reporting policy & 0.50 overall / 0.31 motion & $3.11 \pm 0.02$ & $4.14 \pm 0.15$ \\
\bottomrule
\end{tabular}
\end{adjustbox}
\end{table}

\begin{table}[H]
\centering
\caption{Finite-state reporting policy on PPG-DaLiA. MAE is reported in BPM; coverage is the fraction of windows reported after accept/hold/reject decisions.}
\label{tab:dalia_policy_curve}
\begin{adjustbox}{max width=\columnwidth}
\begin{tabular}{rrrrr}
\toprule
$\lambda_{\mathrm{rej}}$ & Overall coverage & Overall MAE & Motion coverage & Motion MAE \\
\midrule
30 & 1.00 & 7.49 & 1.00 & 12.72 \\
18 & 0.94 & 6.63 & 0.86 & 11.16 \\
12 & 0.62 & 4.24 & 0.41 & 7.24 \\
8  & 0.50 & 3.11 & 0.31 & 4.14 \\
6  & 0.40 & 2.62 & 0.23 & 3.19 \\
4  & 0.30 & 2.26 & 0.17 & 2.48 \\
2  & 0.10 & 1.43 & 0.06 & 1.55 \\
\bottomrule
\end{tabular}
\end{adjustbox}
\end{table}

Sequence decoding lowers full-coverage motion MAE from 14.22 to 12.62~BPM. At 50\% overall coverage, fixed-threshold selective reporting reaches 7.30~BPM motion MAE, while the finite-state policy reaches 4.14~BPM with 0.31 motion coverage. The full policy curve spans full reporting to low-coverage reporting, so each MAE value should be interpreted with its corresponding motion coverage. These results show that the reporting layer improves motion-window output on an independent wrist dataset; direct cross-cohort transfer remains a separate calibration problem.

\section{Summary}
\label{sec:concl_summary}

This thesis studied heart-rate estimation from motion-corrupted wearable PPG. The ring benchmark identifies motion as the main failure mode for standard supervised models. The proposed system separates \emph{estimation} (candidate selection + causal Viterbi) from \emph{reporting} (reliability scoring + learned policy). At 50\% motion-window coverage, selective reporting reduces motion MAE from 10.8 to 6.2~BPM. The estimator fits a microcontroller-class compute budget, and the PPG-DaLiA experiments extend the evaluation beyond the ring cohort. The ablations indicate that the raw-waveform encoder is not needed once candidate features are available, ACC is useful mainly for upstream segment selection, and the reporting policy is most useful at high coverage.

\section{Limitations}
\label{sec:concl_limitations}

\begin{itemize}
    \item Under motion, candidate \textbf{identifiability} (1-of-${\approx}140$ selection, only modestly above chance) is the open bottleneck---and should be measured by chance-corrected top-$k$, not candidate coverage alone.
    \item PPG-DaLiA introduces a different sensor setting (single-channel wrist BVP), so reporting calibration and cross-cohort transfer should be evaluated separately from DaLiA-specific training.
    \item Evaluation is on healthy adults; arrhythmia is untested.
    \item Temporal decoding is not uniformly beneficial. The physiological transition penalty helps most when the reference HR is stable or jumps sharply, but \textbf{over-smooths} windows where the reference changes by a moderate $3$--$5$~BPM between adjacent windows, \emph{increasing} error there (Table~\ref{tab:concl_oversmoothing}). The effect is stable across seeds and transition-penalty settings, and motivates folding reliability into decoding (below).
\end{itemize}

\begin{table}[H]
\centering
\caption{Temporal-decoding gain stratified by the reference-HR change between adjacent windows ($|\Delta y_t|$), aggregated over five seeds. Positive gain means decoding lowers motion-window MAE; the $3$--$5$~BPM bin shows an over-smoothing cost.}
\label{tab:concl_oversmoothing}
\begin{adjustbox}{max width=\columnwidth}
\begin{tabular}{lrrrr}
\toprule
\textbf{$|\Delta y_t|$ (BPM)} & \textbf{$n$} & \textbf{MAE base} & \textbf{MAE Viterbi} & \textbf{Gain} \\
\midrule
$<1$   & 520 & 13.86 & 10.90 & $+2.96$ \\
$1$--$3$ & 245 & \phantom{0}7.41 & \phantom{0}5.21 & $+2.20$ \\
$3$--$5$ & 120 & \phantom{0}8.55 & 11.53 & $\mathbf{-2.98}$ \\
$\ge 5$  & 260 & 18.76 & 17.36 & $+1.40$ \\
\bottomrule
\end{tabular}
\end{adjustbox}
\end{table}

\section{Future Work}
\label{sec:concl_future}

\begin{itemize}
    \item Improve candidate identifiability under motion by reporting chance-corrected top-$k$ selection accuracy and harmonic-confusion rates, in addition to candidate-coverage MAE.
    \item Replace the fixed quadratic transition penalty with a reliability- and motion-adaptive decoder. A Huber or piecewise-linear transition cost is the next decoder variant to test for the $3$--$5$~BPM over-smoothing observed in Table~\ref{tab:concl_oversmoothing}.
    \item Calibrate the reporting layer across cohorts. The reliability score and accept/hold/reject policy should be evaluated with calibration curves and coverage--MAE curves under ring-to-DaLiA transfer, rather than tuning each dataset independently.
    \item Extend evaluation beyond healthy sinus-rhythm recordings. Arrhythmia, low perfusion, and device-placement changes can alter candidate structure, reliability calibration, and hold/reject behavior.
    \item Validate embedded execution on target hardware. The current deployment analysis estimates memory and runtime; future work should measure fixed-point or int8 inference, buffering, and end-to-end latency on the target microcontroller.
\end{itemize}
  \endgroup

\bibliographystyle{aaai}
\bibliography{references,references_extra}

\end{document}